# Rumor Forwarding Prediction Model Based on Uncertain Time Series

*Ruihong Wang*

*Business School, Shandong Normal University,*

*Jinan, Shandong 250014, China*

*Fengming Liu*

*Business School, Shandong Normal University,*

*Jinan, Shandong 250014, China*

The rapid spread of rumors in social media is mainly caused by individual retweets. This paper applies uncertainty time series analysis (UTSA) to analyze a rumor retweeting behavior on Weibo. First, the rumor forwarding is modeled using uncertain time series,including order selection, parameter estimation, residual analysis, uncertainty hypothesis testing and forecast, and the validity of using uncertain time series analysis is further supported by analyzing the characteristics of the residual plot. The experimental results show that the uncertain time series can better predict the next stage of rumor forwarding. The results of the study have important practical significance for rumor management and the management of social media information dissemination.

*Keywords*: Uncertain time series analysis; rumor retweeting; rumor spreading.

## 1.  Introduction

The rise of social media has made the dissemination of information more convenient and widespread, however, it has also provided more opportunities for the spread of false information and rumors. Rumors may not only lead to public misunderstanding, but also cause panic, social instability, and even have far-reaching effects on individuals and society. Therefore, it has become crucial to study and understand the mechanism of rumor spreading and how to predict rumor forwarding.

Many scholars have proposed various rumor prediction algorithms based on the rumor propagation process, aiming to better understand and predict the rumor propagation behavior in social networks. Fang et al.[1] used common classification algorithms to predict whether users retweet rumors based on information dissemination theory; Liu et al.[2] proposed a microblogging network event rumor information diffusion model based on user behavior to calculate the probability of users' retweeting and reading microblogs, so as to predict the dissemination of event information; Jiang et al.[3] proposed an information diffusion prediction model based on the identification of key nodes of the network; Luo et al.[4] proposed a microblog retweet prediction algorithm based on the random forest algorithm for predicting the propagation trend of microblog information; Zhao et al.[5] proposed a retweet count prediction model BCI based on the willingness of retweeting and influence of followers for estimating the scale of information propagation in social networks; Nesi, P.[6] proposed a retweet prediction model based on the evaluation of the tendency of retweeting tweets; Huang et al.[7] classified users' tweets based on Bayesian algorithm to predict their retweeting behavior; Morchid et al.[8] predicted information retweets based on tweet features to reveal the trend of information dissemination in social networks.

Time series analysis is a powerful tool that can help us capture the dynamic nature and uncertainty in the process of information dissemination.Ye et al.[9] analyzed and predicted the evolution of confirmed COVID-19 cases in China by proposing an uncertainty time series model; He et al.[10] extended the well-known skyline analysis method to uncertain time series and proposed an effective method to solve the problem of probabilistic skyline computation for uncertain time series; Ye et al.[11] modeled and predicted the birth rate of China's population using uncertain time series analysis; Ye [12] modeled and predicted China's grain yield using uncertain time series analysis.

This paper proposes a method based on uncertain time series analysis to model and predict the number of rumor retweets. First, uncertain time series analysis is performed on the number of retweets of a rumor to obtain an uncertain time series model without outliers, and by analyzing the characteristics of the residual plots, we explain why uncertain statistics are used instead of probability statistics. Finally, the application of uncertain sequence analysis to rumor propagation is summarized based on the model results.

## 2. Uncertain time series analysis

### 2.1. Uncertain autoregressive model

Suppose $X_t$ is an observation at time $t$, $t = 1, 2, ..., n$. Then the sequence of observations

$$X_1, X_2, ..., X_n \tag{1}$$

is called a time series. To model the time series (1), Yang and Liu [13] proposed an uncertain autoregressive model,

$$X_t = a_0 + \sum_{i=1}^{k} a_i X_{t-i} + \varepsilon \tag{2}$$

where k is called the order of the uncertain autoregressive model, which determines how many past observations are considered in the model; $a_0, a_1, ..., a_k$ are unknown parameters, which needs to be estimated from the data; and $\varepsilon$ is the uncertain disturbance term(uncertain variable).

### 2.2. Parameter estimation

Based on the uncertain time series $X_1, X_2, ..., X_n$, the least squares estimate of $(a_0, a_1, ..., a_k)$ in the uncertain autoregressive model (2) is the solution to the minimization problem

$$\min_{a_0, a_1, ... a_k} \sum_{t=k+1}^{n} \left( X_t - a_0 - \sum_{i=1}^{k} a_i X_{t-i} \right)^2.$$

If the minimum solution is $\left( \hat{a}_0, \hat{a}_1, ..., \hat{a}_k \right)$, then the fitted autoregressive model is

$$X_t = \hat{a}_0 + \sum_{i=1}^{k} \hat{a}_i X_{t-i}$$

### 2.3. Residual analysis

For each index $t(t = k+1, k+2, ..., n)$, Yang and Liu[13] proposed that the difference between the actual observations and the model predictions

$$\hat{\varepsilon}_t = X_t - \hat{a}_0 - \sum_{i=1}^{k} \hat{a}_i X_{t-i} \tag{3}$$

is called the $t$th residual. The residual $\varepsilon_{k+1}, \varepsilon_{k+2}, \ldots, \varepsilon_n$ will be regarded as a sample of the uncertain disturbance term $\varepsilon$ in model (2). Therefore, the expected value $\hat{e}$ and variance $\hat{\sigma}^2$ of the uncertain disturbance term $\varepsilon$ can be estimated as

$$\hat{e} = \frac{1}{n-k} \sum_{t=k+1}^{n} \varepsilon_t$$

and

$$\hat{\sigma}^2 = \frac{1}{n-k} \sum_{t=k+1}^{n} \left(\varepsilon_t - \hat{e}\right)^2,$$

respectively. Thus, the estimated disturbance term $\hat{\varepsilon}$ is an uncertain variable with expectation $\hat{e}$ and variance $\hat{\sigma}^2$. If we further assume that $\hat{\varepsilon}$ obeys the normal uncertainty distribution, we obtain an uncertain autoregressive model

$$X_t = \hat{a}_0 + \sum_{i=1}^{k} \hat{a}_i X_{t-i} + \mathrm{N}\left(\hat{e}, \hat{\sigma}\right) \tag{4}$$

### 2.4. Uncertain hypothesis test

Based on the time series $X_1, X_2, \ldots X_n$, we can get the uncertain autoregressive model as

$$X_t = \hat{a}_0 + \sum_{i=1}^{k} \hat{a}_i X_{t-i} + \mathrm{N}\left(\hat{e}, \hat{\sigma}\right) \tag{5}$$

In order to test whether the uncertain autoregressive model (5) fits the observed data well, we should test whether the normal uncertainty distribution $\mathrm{N}\left(\hat{e}, \hat{\sigma}\right)$ fits the $n-k$ residuals $\varepsilon_{k+1}, \varepsilon_{k+2}, \ldots, \varepsilon_n$ determined by Equation (3), i.e.

$$\varepsilon_{k+1}, \varepsilon_{k+2}, \ldots, \varepsilon_n \sim \mathrm{N}\left(\hat{e}, \hat{\sigma}\right).$$

Given the significance level $\alpha$ (e.g. $\alpha = 0.05$), the test (Ref. [14]) is

$W = \{(z_{k+1}, z_{k+2}, \ldots, z_n) : \text{there are more than } \alpha \text{ of indexes } t\text{'s with } k+1 \leq t \leq n$

such that $z_t < \phi^{-1}\left(\dfrac{\alpha}{2}\right)$ or $z_t > \phi^{-1}\left(1-\dfrac{\alpha}{2}\right)\right\}$ ,

where

$$\phi^{-1}(\alpha) = \hat{e} + \dfrac{\hat{\sigma}\sqrt{3}}{\pi}\ln\dfrac{\alpha}{1-\alpha}.$$

If the vector of $n-k$ residuals $\varepsilon_{k+1}, \varepsilon_{k+2}, \ldots, \varepsilon_n$ belongs to $W$, i.e.

$$(\varepsilon_{k+1}, \varepsilon_{k+2}, \ldots, \varepsilon_n) \in W ,$$

then the uncertain autoregressive model (5) is not a good fit to the observed data. In this case, we have to choose an uncertain time series model again. If $(\varepsilon_{k+1}, \varepsilon_{k+2}, \ldots, \varepsilon_n) \notin W$, then the uncertain autoregressive model (5) is a good fit to the observed data. In both cases, determining whether the residual vector falls within a particular subspace can help us assess the fit of the model to determine whether the model is appropriate to describe the observed data.

### 2.5. *Forecast uncertain variable*

Based on the estimated uncertain autoregressive model (4), Yang and Liu[13] proposed that for the time series $X_1, X_2, \ldots, X_n$ the predicted uncertain variable $X_{n+1}$ is

$$\hat{X}_{n+1} = \hat{a}_0 + \sum_{i=1}^{k} \hat{a}_i X_{n+1-i} + \mathrm{N}\left(\hat{e}, \hat{\sigma}\right) \qquad (6)$$

### 2.6. *Forecast value*

Based on the uncertain variables of prediction (6), Yang and Liu[12] proposed to define the predicted value as the expected value of $\hat{X}_{n+1}$, i.e.

$$\hat{\mu} = \hat{a}_0 + \sum_{i=1}^{k} \hat{a}_i X_{n+1-i} + \hat{e} \qquad (7)$$

### 2.7. *Confidence interval*

According to (6),(7) and the algorithm of uncertain variables, we can get the predicted uncertain variable $\hat{X}_{n+1}$ obeys the normal uncertainty distribution $\mathrm{N}\left(\hat{\mu}, \hat{\sigma}\right)$, i.e.,

$$\hat{\psi}(z) = \left(1 + \exp\left(\dfrac{\pi\left(\hat{\mu}-z\right)}{\sqrt{3}\hat{\sigma}}\right)\right)^{-1}$$

Taking $\alpha$ (e.g. 95%) as a confidence level, it is easy to verify that the minimum interval $[a,b]$ of $\hat{\psi}(b) - \hat{\psi}(a) \geq \alpha$ is

$$\left[\hat{\mu} - \frac{\hat{\sigma}\sqrt{3}}{\pi}\ln\frac{1+\alpha}{1-\alpha}, \hat{\mu} + \frac{\hat{\sigma}\sqrt{3}}{\pi}\ln\frac{1+\alpha}{1-\alpha}\right].$$

Due to

$$M\left\{a \leq \hat{X}_{n+1} \leq b\right\} \geq \hat{\psi}(b) - \hat{\psi}(a) = \alpha,$$

Yang and Liu [12] proposed the $\alpha$ confidence interval for $X_{n+1}$ as

$$\hat{\mu} \pm \frac{\hat{\sigma}\sqrt{3}}{\pi}\ln\frac{1+\alpha}{1-\alpha}.$$

3. **Analysis of the number of rumor retweets for uncertain time series**

In this study, we take a typical case of online rumor dissemination as an example, which involves the incident of "the deputy director of Wuhan Health Commission fled to a mansion in Shanghai after infection". We analyzed the trend of rumor retweets by crawling and organizing data from microblogs related to the event. The time range of data crawling was from 14:00 pm to 2:00 am the next day, totaling 12 hours, and a total of 1,387 retweeted microblog data were collected. Through detailed analysis and fitting of these data, we can more accurately reflect the propagation pattern of the rumor and provide support for taking corresponding management and control measures. The specific crawled data are detailed in Table 1 and Figure 1, and they will help us gain insight into the rumor spreading in order to better understand and deal with similar events.

| 78 | 175 | 257 | 334 | 392 | 463 | 529 | 582 | 628 | 667 |
|---|---|---|---|---|---|---|---|---|---|
| 701 | 724 | 749 | 788 | 822 | 868 | 899 | 933 | 965 | 1000 |
| 1021 | 1048 | 1072 | 1088 | 1112 | 1123 | 1132 | 1147 | 1158 | 1170 |
| 1186 | 1197 | 1210 | 1222 | 1235 | 1243 | 1252 | 1259 | 1267 | 1279 |
| 1285 | 1291 | 1295 | 1299 | 1307 | 1314 | 1321 | 1331 | 1334 | 1336 |
| 1339 | 1346 | 1351 | 1354 | 1356 | 1357 | 1359 | 1360 | 1363 | 1368 |
| 1371 | 1371 | 1372 | 1374 | 1376 | 1378 | 1379 | 1380 | 1380 | 1380 |
| 1386 | 1387 | | | | | | | | |

Table 1

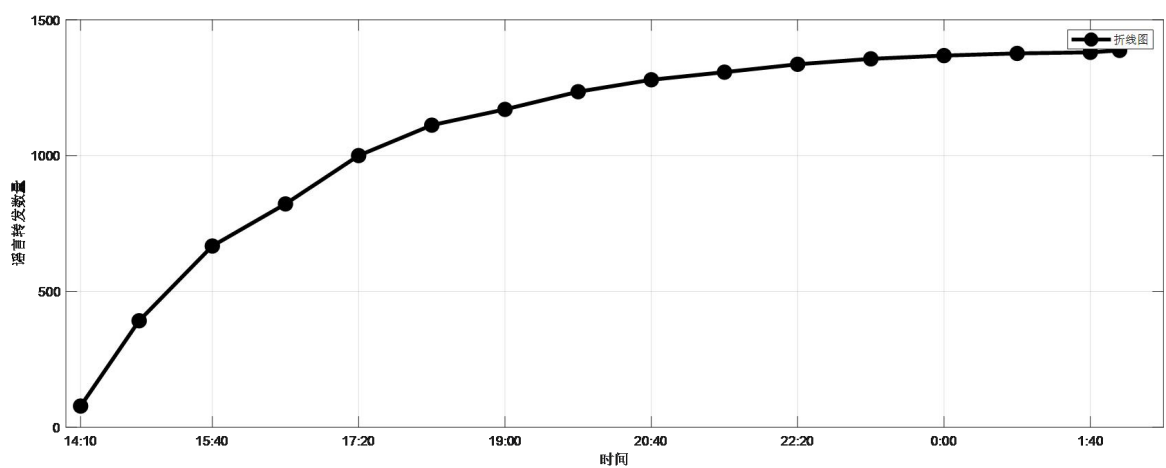

Figure 1

We will apply the method of uncertain time series analysis to model the number of rumor forwards. Denote the time as t = 14:10, 14:30, ..., 2:00 covering the time range from 14:00 to 2:00. The data observed in Table 1 is represented in the following way:

$$X_t, t = 14:10, 14:20, ..., 2:00, \qquad (8)$$

where $X_t$ are the number of rumor retweets at time $t$, $t = 14:10, 14:20, ..., 2:00$, respectively. To model the number of rumor retweets, we used an uncertain autoregressive model,

$$X_t = a_0 + \sum_{i=1}^{k} a_i X_{t-i} + \varepsilon \qquad (9)$$

where $k$ is the order to be determined, $a_0, a_1, ..., a_k$ are unknown parameters to be estimated, and $\varepsilon$ is the uncertainty perturbation term.

### 3.1. *Determination order of uncertain autoregressive model*

In order to determine the optimal order $k$ in the uncertain autoregressive model (9), we adopt the rolling window cross-validation method proposed by Liu and Yang[15]. The maximum order is set to 10, and the length of the training set is set to 22. With this method, the average testing errors(ATEs) corresponding to different orders can be calculated, as shown in Fig. 2.

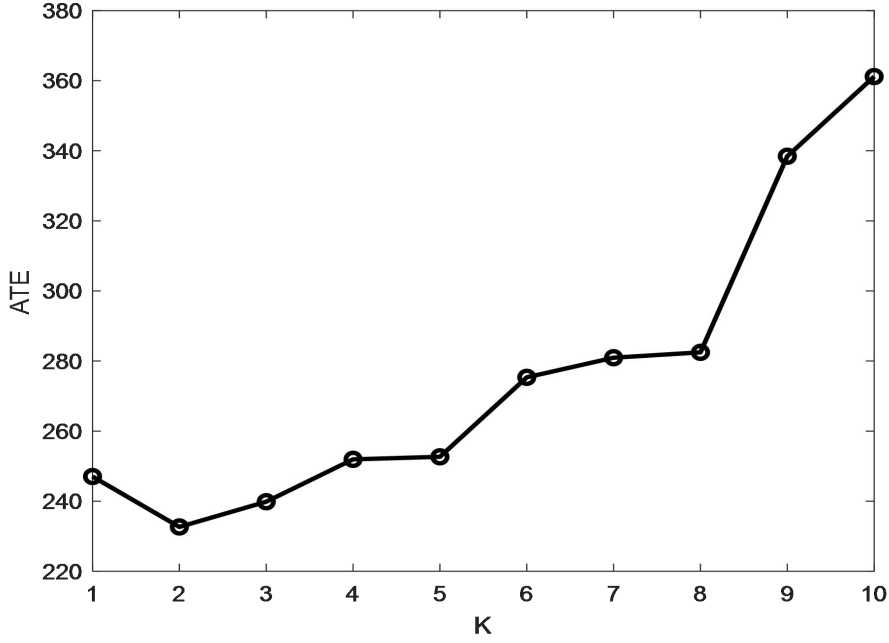

Fig. 2 Average Test Error (ATE) for cross validation

As shown in Fig. 2, we choose the order $k = 2$, since ATE (2) is the minimum value. Therefore, we use the following uncertain autoregressive model:

$$X_t = a_0 + a_1 X_{t-1} + a_2 X_{t-2} + \varepsilon \tag{10}$$

where $a_0$, $a_1$ and $a_2$ are the unknown parameters to be estimated, and $\varepsilon$ is the uncertainty perturbation term.

### 3.2. *Parameter estimation*

Based on the observed data(8), the minimization problem was solved using MATLAB:

$$\min_{a_0, a_1, a_2} \sum_{t=14:00}^{2:00} (X_t - a_0 - a_1 X_{t-1} - a_2 X_{t-2})^2,$$

The minimum estimates of $a_0$, $a_1$ and $a_2$ are obtained as

$$\hat{a}_0 = 50.9313, \hat{a}_1 = 1.3276, \hat{a}_2 = -0.3641$$

Therefore, the fitted autoregressive model is

$$X_t = 50.9313 + 1.3276 X_{t-1} - 0.3641 X_{t-2} \tag{11}$$

### 3.3. *Residual analysis*

Based on the fitted autoregressive model (11), from

$$\varepsilon_t = X_t - 50.9313 - 1.3276 X_{t-1} - 0.3641 X_{t-2}$$

we can get 70 residuals as shown in Figure 3.

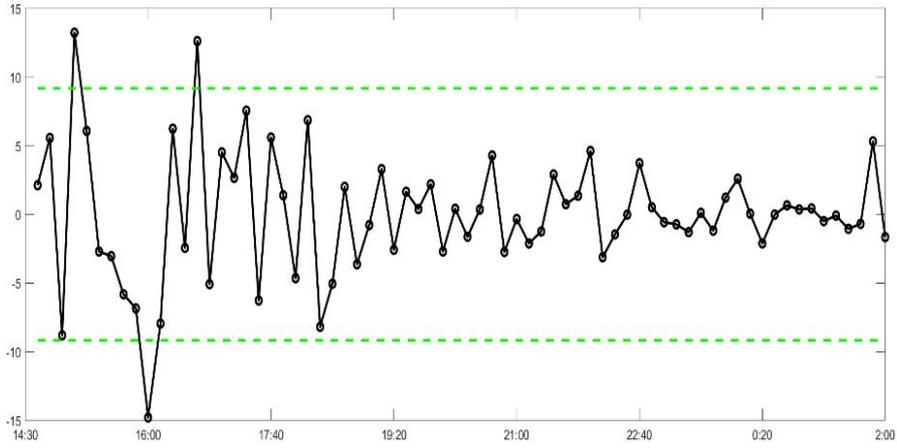

Figure 3. Residual plots for uncertain autoregressive models

In model (10), if the perturbation term is considered as a random variable, the residuals in Fig. 3 should belong to the same probability distribution population according to the probabilistic view. However, when we use the two-sample Kolmogorov-Smirnov test to compare whether the two parts of the 70 residuals,

$$\left(\varepsilon_{14:10}, \varepsilon_{14:20}, \ldots, \varepsilon_{18:50}\right) \text{ and } \left(\varepsilon_{17:00}, \varepsilon_{17:10}, \ldots, \varepsilon_{2:00}\right)$$

come from the same population, we get a p-value of less than 0.05 by using the function "kstest 2" in Matlab. this means that the residuals are neither from the same population, nor white noise, nor a random sequence in the probabilistic sense. In this case, if we still use probabilistic time series analysis, the perturbation terms of the estimated distribution function will not be close enough to the true frequencies (residuals). This is why we treat the perturbation term as an uncertain variable rather than a random variable, and why we chose uncertain time series analysis over probabilistic time series analysis in this example. In short, if the estimated distribution function is close enough to the actual frequency distribution, then probabilistic time series analysis may be more appropriate. However, in our case, uncertain time series analysis seems to be more consistent with the nature and distribution of the residuals.

Then the expected value of the uncertain disturbance term $\varepsilon$ is estimated as

$$\hat{e} = \frac{1}{70} \sum_{t=14:10}^{2:00} \varepsilon_t = 0.0000$$

and the variance is estimated as

$$\hat{\sigma}^2 = \frac{1}{55} \sum_{t=14:10}^{2:00} (\varepsilon_t - \hat{e})^2 = 4.5437^2 .$$

Thus, we obtain an uncertain autoregressive model

$$X_t = 50.9313 + 1.3276 X_{t-1} - 0.3641 X_{t-2} + N(0, 4.5437) \qquad (12)$$

### 3.4. *Uncertain hypothesis test*

Finally, we will use uncertainty hypothesis testing to assess whether the uncertainty autoregressive model (12) fits the observed data well. Specifically, we will test whether the normal uncertainty distribution $N(0, 4.5437)$ fits the residuals $\varepsilon_{14:10}, \varepsilon_{14:20}, ..., \varepsilon_{2:00}$. Given a significance level $\alpha = 0.05$, the test from $\alpha \times 70 = 3.5$ that the test is

$$W = \{(z_{14:10}, z_{14:10}, ..., z_{2:00}) : \text{there are at least four indexes } t's \text{ with}$$

$$14:10 \leq t \leq 2:00 \text{ such that } z_t < -9.1774 \text{ or } z_t > 9.1774\}$$

Since there is only $\varepsilon_{15:00}, \varepsilon_{16:00}, \varepsilon_{16:40} \notin [-9.1774, 9.1774]$, we have $(\varepsilon_{14:10}, \varepsilon_{14:20}, ..., \varepsilon_{2:00}) \notin W$.

Therefore, based on the test results, we can conclude that the uncertain autoregressive model (12) is well fitted to the observed data. This test result shows that the model's uncertainty assumptions are consistent with the observed data, further supporting the applicability of the uncertainty autoregressive model.

### 3.5. *Forecast*

Based on the uncertain autoregressive model (12), the uncertain variables for predicting the number of rumor forwards at 24 hours is

$$\hat{X}_{2:10} = 50.9313 + 1.3276 \times 1387 - 0.3641 \times 1386 + N(0, 4.5437),$$

i.e., $\hat{X}_{24} \sim N(1388, 4.5437)$ which in turn predicts the number of rumor retweets to be 1388, and the 95% confidence interval is

$$1388 \pm \frac{4.5437\sqrt{3}}{\pi} \ln \frac{1+0.95}{1-0.95},$$

i.e., $1388 \pm 9.1774 = [1378.8226, 1397.1774]$.

### 4. Revelations

First, through the in-depth analysis of uncertain time series, we can more accurately predict the trend and pattern of the number of rumor retweets. This provides a new way for rumor management, enabling managers to take more timely measures to limit the spread of rumors and thus reduce their social impact.

Second, the application of uncertain time series makes monitoring rumor spreading more real-time and precise. This means that administrators can more quickly identify trends in the spread of rumors and intervene accordingly, for example by publishing truthful information or shutting down sources of false information.

Further, analysis based on uncertain time series can help policymakers allocate resources more efficiently in response to spikes in rumor spread. This can help to improve the efficiency of rumor management with limited resources.

Finally, the study of uncertain time series contributes to an in-depth understanding of the mechanisms and influencing factors of rumor propagation. Through the detailed analysis of time series, we are able to identify the key nodes and trends of rumor propagation, thus gaining a more comprehensive understanding of the dynamics of information dissemination on social media. This provides strong support for the development of more effective rumor management strategies.

### 5. Conclusion

Uncertainty time series analysis is based on the theory of uncertainty and predicts future values from previously observed data. In this paper, we propose a new method for modeling and predicting the number of rumor retweets based on uncertainty time series analysis. The experimental results show that the method can predict the number of rumor retweets more accurately and can better deal with the uncertainty in rumor propagation; secondly, this paper also explains why uncertain time series analysis is used instead of probabilistic time series analysis, pointing out that the residuals neither come from the same general population, nor are they white noise, nor are they random sequences in the probabilistic sense. In future research, the application of uncertain time series in rumor management can be further explored in terms of model improvement, data source integration, and integration with other analytical methods. These studies will help to further improve the effectiveness of rumor governance. In conclusion, uncertain time series analysis provides new tools and insights for rumor governance, which is expected to help us better understand and respond to the problem of rumor spreading on social media.

**Acknowledgement**

**This research was supported in part by the National Social Science Foundation of China (No. 21BGL001), the National Natural Science Foundation of China(71701115), Shandong Natural Science Foundation (ZR2020MG003), Special Project for Internet Development of Social Science Planning Special Program of Shandong Province (17CHLJ23)**